\documentstyle[11pt,twoside]{article}
\newcount\Mac  \Mac=2  
\newcommand{\ifMac}[2]{\ifnum\Mac=1 #1 \else #2 \fi}

\newcommand{\GeV}{\,{\rm GeV}}
\newcommand{\TeV}{\,{\rm TeV}}

\newcommand{\journal}[4]{{\em #1 \bf #2} (#3) #4}
\newcommand{\NP}[3] {\journal{Nucl. Phys.}{#1}{#2}{#3}}
\newcommand{\PRL}[3]{\journal{Phys. Rev. Lett.}{#1}{#2}{#3}}
\newcommand{\PL}[3] {\journal{Phys. Lett.}{#1}{#2}{#3}}
\newcommand{\PR}[3] {\journal{Phys. Rev.}{#1}{#2}{#3}}
\newcommand{\mhu}{m_{h^{\rm u}}}
\newcommand{\mhd}{m_{h^{\rm d}}}

\newcommand{\meR}{m_{\tilde{e}_R}}

\newcommand{\eR}{\tilde{e}_R}
\newcommand{\eq}[1]{~{\rm (\ref{eq:#1})}}
\newcommand{\md}[1]{\langle#1\rangle}
\newcommand{\MGUT}{M_{\rm GUT}}
\def\Red{}
\def\Black{}
\def\Blue{}

\newcommand{\lascia}[1]{}
\makeatletter
\newcounter{alphaequation}[equation]
\def\thealphaequation{\theequation\hbox to
0.6em{\hfil\alph{alphaequation}\hfil}}
\def\eqnsystem#1{
\def\@eqnnum{{\rm (\thealphaequation)}}
\def\@@eqncr{\let\@tempa\relax \ifcase\@eqcnt \def\@tempa{& & &} \or
  \def\@tempa{& &}\or \def\@tempa{&}\fi\@tempa
  \if@eqnsw\@eqnnum\refstepcounter{alphaequation}\fi
\global\@eqnswtrue\global\@eqcnt=0\cr}
\refstepcounter{equation} \let\@currentlabel\theequation \def\@tempb{#1}
\ifx\@tempb\empty\else\label{#1}\fi
\refstepcounter{alphaequation}
\let\@currentlabel\thealphaequation
\global\@eqnswtrue\global\@eqcnt=0 \tabskip\@centering\let\\=\@eqncr
$$\halign to \displaywidth\bgroup \@eqnsel\hskip\@centering
$\displaystyle\tabskip\z@{##}$&\global\@eqcnt\@ne
\hskip2\arraycolsep\hfil${##}$\hfil& \global\@eqcnt\tw@\hskip2\arraycolsep
$\displaystyle\tabskip\z@{##}$\hfil
\tabskip\@centering&\llap{##}\tabskip\z@\cr}

\def\endeqnsystem{\@@eqncr\egroup$$\global\@ignoretrue} \makeatother

  \def\Ord{{\cal O}}    
\def\circa#1{\,\raise.3ex\hbox{$#1$\kern-.75em\lower1ex\hbox{$\sim$}}\,}

\begin{document}

\parbox{10em}{\raggedleft \em 1 November 1996\\  \bf hep-ph/9611204}\hfill
\parbox{10em}{\bf UAB--FT--403 \\ \bf FT--UAM 96/43}\\[5mm]

\centerline{\huge\bf\Red Naturalness upper bounds on}
\centerline{\huge\bf      gauge mediated soft terms}
\bigskip\bigskip\Black
\centerline{\large\bf Paolo Ciafaloni}\smallskip
\centerline{\large\em IFAE - Grup de F\'{\i}sica Te\`orica, Edifici Cn,
Universitat  Aut\`onoma}
\centerline{\large\em de Barcelona,
08193 Bellaterra, Espa\~{n}a and INFN -- Frascati -- Italy}
\medskip
\centerline{\large and \bf Alessandro Strumia}\smallskip
\centerline{\large\em Departamento de F\'{\i}sica Te\'orica,
Universidad Aut\'onoma de Madrid,}
\centerline{\large\em 28049, Madrid, Espa\~na {\rm and}
INFN, Sezione di Pisa, I-56126 Pisa, Italia}
\bigskip\bigskip\Blue \centerline{\large\bf Abstract}
\begin{quote}\large\indent
After a general discussion about the quantitative meaning
of the naturalness upper bounds on
the masses of supersymmetric particles,
we compute these bounds in models with gauge-mediated soft terms.
We find interesting upper limits on the right-handed slepton masses that,
unless the messenger fields are very light, 
disfavor minimal models with large messenger content.
Deep unphysical minima, that however
turn out to be not dangerous, are usually present in such models.
The $\mu$-problem can be solved by adding a light singlet
only at the price of a large amount of fine tuning
that gives also rise to heavy sparticles and large $\tan\beta$.
\end{quote}\Black

\section{Introduction}
In supersymmetric theories the $Z$-boson mass
is determined as a function of the soft terms
in the Higgs potential.
For this reason it is unnatural
that these soft terms be much larger than $M_Z$.
In specific models that try to understand the
origin of the soft terms,
specific relations hold between the sfermion masses
and the Higgs soft terms,
allowing to
place specific naturalness bounds on sparticle masses.
The case of supergravity mediated soft terms~\cite{SuGraSoft}
has been extensively studied under the assumptions of
universal boundary conditions~\cite{FT1,FT2,FTmedio}
or with relations suggested by unification theories~\cite{FTgut,FT*2}.
In this paper we will study the case
of gauge-mediated soft terms~\cite{GaugeSoft}.
If soft terms are mediated by ordinary gauge interactions,
the combination of Higgs soft mass terms
relevant for the determination of $M_Z$
is tipically much larger than the small
soft mass terms of the right-handed sleptons, $\tilde{e}_R$,
that must therefore be particularly light for naturalness reasons.
Even though some crucial ingredients of the Higgs potential,
the $\mu$ and $B\cdot\mu$ terms,
cannot be mediated by gauge interactions~\cite{GMmu},
we discuss how, under reasonable assumptions, it is nevertheless
possible to compute and quantify the naturalness bounds.
Our result is that in minimal models, unless
cancellations between the known and the unknown missing
contributions of more than one order of magnitude are present,
the right-handed sleptons are lighter than $100\GeV$.
Although possible, such large cancellations have a
small probability of about 10\%.

In section~\ref{FT} we reconsider the general problem
of quantifying the naturalness bounds in supersymmetric theories.
In section~\ref{GM} we compute the naturalness
bounds in a wide class of minimal gauge mediation scenarios
with the MSSM as the low energy theory.
We also present in appendix~A the exact solutions
to the one-loop renormalization group equations (RGEs).
In section~\ref{NMSSM} we focus on the models (`NMSSM') where
the $\mu$ problem is solved by adding a light singlet
to the observable MSSM fields.
We find that, with gauge-mediated soft terms,
the actual minimum of the potential is unphysical
unless some parameters are unnaturally fine-tuned in such a way
that also results in heavy sparticles and large $\tan\beta$.

\section{Naturalness bounds}\label{FT}
In supersymmetric models the $Z$ boson mass
is obtained, by minimizing the potential,
in function of the soft terms that enter in the Higgs potential.
For example, in the MSSM at tree level, one has
\begin{equation}\label{eq:MZ^2}
M_Z^2=-2|\mu|^2 + 2\frac{\mhd^2-\mhu^2\tan^2\beta}{\tan^2\beta-1}
\end{equation}
where $\mu$ is the `$\mu$-term' and
$\mhu^2$ ($\mhd^2)$ is  the
soft mass  of the Higgs fields
$h_{\rm u}$ ($h_{\rm d}$) that gives mass to
up quarks (down quarks and leptons).
Since there are contributions of both signs it is possible,
although unnatural, that the sparticles be  much heavier
than the $Z$ boson due to accidental cancellations between
different terms in eq.\eq{MZ^2}.
Naturalness considerations imply upper bounds on the sparticle masses, but
it is difficult to give them an unambiguous quantitative meaning.

Naturalness bounds had originally been quantified~\cite{FT1} by constraining
the sensitivity of $M_Z^2$ with respect to variations
of a set of parameters $\{\wp\}$ (soft terms, gauge and Yukawa couplings)
chosen to be the `fundamental' ones.
More in detail,
it was required that the fractional variation of $M_Z$ (or $M_Z^2$)
with respect to fractional variations of the parameters $\wp$
\begin{equation}\label{eq:FT}
\Delta[p] \equiv \frac{\wp}{M_Z^2}\frac{\partial M_Z^2}{\partial\wp},
\qquad
\Delta \equiv \max_\wp \Delta[\wp]
\end{equation}
be smaller than some maximum allowed value,
$\Delta<\Delta_{\rm lim}$,
originally chosen as $10$.
Following the lines of ref.~\cite{FT2,FTmedio}, where this definition
has been criticized, let us now show some
examples where it turns out to be too restrictive or inadequate.
\begin{itemize}
\item[i.]
Let us first try to apply it to the
case in which the supersymmetry breaking scale,
and consequently the $Z$-boson mass, are dynamically determined,
for example through gaugino condensation in a `hidden' sector.
In such a case
$$M_Z \approx M_{\rm Pl} \cdot e^{-c/g^2_H},$$
where $g_H$ is the hidden sector gauge coupling constant
renormalized at $M_{\rm Pl}$
and $c$ is a numerical constant,
dependent on
the one-loop coefficient of the hidden
gauge group $\beta$-function
and on the mediation scheme.
Even though $\Delta[g_H]\sim \ln M_{\rm Pl}/M_Z$
is quite large, there is nothing unnatural.

\item[ii.]
As a second more practical
example, let us   consider the sensitivity of the
$Z$-boson mass with respect to
the top quark Yukawa coupling, $\lambda_t$.
Since the pole top mass and
the soft terms of the fields $h_{\rm u}$, $\tilde Q$ and $\tilde t_R$
involved in $\lambda_t$ all run towards an infrared fixed point, 
they are very weak functions of $\lambda_t(\MGUT)$.
The fine-tuning can thus be significant
if we choose $M_t$
as a free parameter, even when the soft terms are of $\Ord(M_Z^2)$,
but becomes very mild if we choose $\lambda_t$ renormalized
at an high scale.
More generally, definition\eq{FT} depends
on the parametrization of the parameter space
in a possibly significant way.
\end{itemize}
In ref.~\cite{FTmedio} these
criticisms have been taken into account
by substituting the sensitivity $\Delta$ of eq.\eq{FT} with
a differently normalized one,
$\Delta/\md{\Delta}$ obtained dividing $\Delta$
by a suitably defined mean sensitivity $\md{\Delta}$.
Somewhat less restrictive bounds have been obtained in this way.
Since the numerical difference is not totally irrelevant,
we will discuss such issues,
choosing a simpler criterion based on solid arguments that,
in some relevant case,
turns out to be numerically equivalent to the original
one~\cite{FT1}. We will now show that
{\em there is an unnaturally small probability $p\approx\Delta^{-1}$ that
the single `contributions' to $M_Z^2$ in eq.\eq{MZ^2} be $\Delta$ times
bigger than their sum, $M_Z^2$.}

Of course the difference between the `interesting' case
of eq.\eq{MZ^2} and the one of example~i.\
is the possibility of cancellations between different contributions.
As a consequence, while in example i.\ small variations of the parameters
can only render $M_Z^2$ orders of magnitude smaller or larger,
in the case of interest small variations of the parameters
can make $M_Z^2$ negative (or more exactly,
erase the SM-like minimum when
the particular point $M_Z^2=0$ is reached),
leading to an undesired qualitative change in the physics.
This simple fact is at the basis of the difference between the two cases:
it is only in the actually interesting case, eq.\eq{MZ^2},
that imposing an  {\em upper limit} on $M_Z$ forces the parameter space
to shrink to a small region.
In example i. instead, an upper limit on the $Z$-boson mass,
$M_Z<M_Z^{\rm lim}$,
is satisfied in a large part of the parameter space $g_H< g_H^{\rm lim}$.
Only in the first case a too light $Z$ boson is unnatural.
This difference between
the two cases cannot be seen if one only looks at
the derivatives of $M_Z^2$,
that determine the fine-tuning as in eq.\eq{FT}.
In fact, even though in both cases $M_Z^2$ has a
strong dependence on $\wp$, it is
only in the `actual' case,
where it is possible for $M_Z^2$ to vanish,
that a first order Taylor expansion of $M_Z^2(\wp+\delta\wp)$
gives a correct approximation of the real behavior.
In this case, a $Z$ boson much lighter than the soft terms,
as can be obtained for some value $\wp$ of the parameters,
is characteristic of only a
small range $\wp-\delta\wp\ldots \wp$, where
$\delta\wp \approx \wp/\Delta[\wp]$.
We now have to quantify the
probability that the parameters $\wp$ lie in a given small range.

In the case that the value of some
particular parameter $\wp$ 
(for example $\lambda_t$, or $g_3$, or $g_{\rm GUT}$)
is known with some error $\sigma_\wp$,
the probability that $\wp$ lies in some small range $\delta \wp$,
compatible with the measured value, is $p\sim \delta \wp/\sigma_\wp=
\Delta[\wp]^{-1}\cdot(\wp/\sigma_\wp)$.
Since $M_t$ and $g_3$ are today known
with a $\sim$5\% error the corresponding naturalness
bounds are not the most interesting ones.
Even though, for this reason, we have not presented
the obvious precise expression of $p$ in terms
of the probability distribution of the measured parameters,
it is useful to remark that such definition
evades also the criticism discussed in ii.
In simpler terms, this happens because
the variation of $M_Z^2$ is the same
when $M_t$ takes its extremely allowed values,
or when $\lambda_t(\MGUT)$ takes its corresponding ones.

Let us now consider the opposite case of
parameters that have not been measured, such as the soft terms.
Since we have no reason for asserting that the particular values
$\wp-\delta\wp\ldots\wp$ necessary to get the desired cancellation
are more probable than, say, any value less than $\wp$,
we must use $\sigma_\wp\sim\wp$.
The probability $p$ of getting an accidental
cancellation more precise than $\Delta[\wp]^{-1}$ is thus
$p \sim \delta \wp/\sigma_\wp=\Delta[\wp]^{-1}$.
In other words the choice of a limiting allowed value
of the sensitivity $\Delta[\wp]$ is nothing more
than a choice of a `confidence limit'
on unprobable cancellations.
To conclude, we choose to normalize the probability so that it is 1
in situations in which we see nothing of unnatural, namely
when no accidental cancellation is present.

\smallskip

We have thus justified the criterion that we employ.
For definiteness we will now briefly exemplificate it,
computing the corresponding
bounds in the
case of the `MSSM with universal soft terms
at the unification scale'\footnote{A more complicate and realistic
example can be found in ref.~\cite{FT*2}.}.
In this case we obtain, using the RGE-improved
tree-level potential, and
for $\lambda_t$ values near to its infrared fixed point
\begin{equation}\label{eq:MZ^2SuGra}
M_Z^2=-2|\mu|^2 + \frac{2 + \Ord(1)\tan^2\beta}{\tan^2\beta-1}m_0^2 +
\frac{\Ord(1.5) + \Ord(10) \tan^2\beta}{\tan^2\beta-1}M_2^2.
\end{equation}
The dependence on $A_0$ is negligible.
Imposing $\Delta<\Delta_{\rm lim}$ we get\footnote{We have considered the variations
of $M_Z^2$ at fixed $\tan\beta$;
other authors prefer to keep fixed the `$B$-parameter' of the Higgs potential,
obtaining an higher value of $\Delta$.
The difference is however not much significative,
provided that $M_Z^2$ is extracted from the
minimization of the full one-loop potential, as we do in this article.},
for moderate values of $\tan\beta\sim 2$,
\begin{equation}\label{eq:SuGraFT}
|\mu|\circa{<} M_Z\sqrt{\frac{\Delta_{\rm lim}}{2}},\qquad
m_0^2\circa{<} M_Z^2\frac{\Delta_{\rm lim}}{2},\qquad
M_2\circa{<} M_Z\sqrt{\frac{\Delta_{\rm lim}}{10}}
\end{equation}
The corresponding bounds on the sparticle masses
can be easily derived from their
expression in terms of $\mu$, $m_0$ and $M_2$~\cite{RGEinMSSM},
and from eq.\eq{MZ^2SuGra}, that connects the various soft parameters.

We now want to apply these considerations to the
more predictive case
of soft terms mediated by MSSM gauge interaction.

\begin{figure}[t]\setlength{\unitlength}{1cm}
\begin{center}\begin{picture}(15.5,9)
\ifMac
{\put(0,0){\special{picture eRN}}}
{\put(0,1){\includegraphics{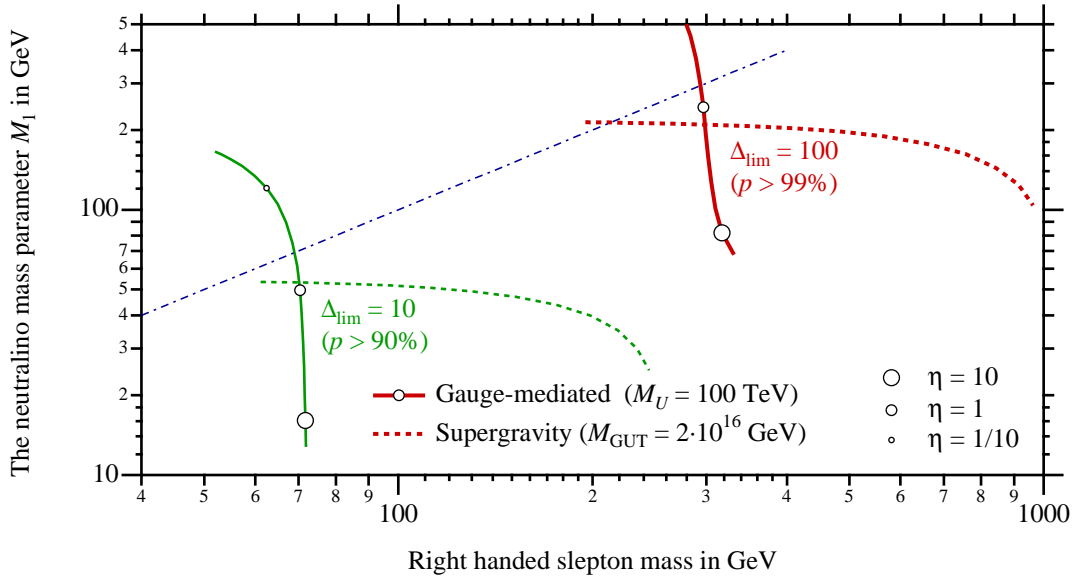}}}
\end{picture}
\caption[fig1]{\em Naturalness upper bounds in the plane $(M_1,\meR)$ with
$\tan\beta=2$
in the cases of universal supergravity-mediated
and of gauge-mediated
soft terms in models with different values of $\eta$, defined
in eq.\eq{GM}.
Above (below) the straight line
a slepton (a neutralino) is the lightest super-partner.\label{fig:eRN}}
\end{center}
\end{figure}

\begin{figure}[t]\setlength{\unitlength}{1cm}
\begin{center}
\begin{picture}(16,8)
\ifMac
{\put(-1,0){\special{picture tanB}}
 \put(7.8,0){\special{picture Eta}}}
{\put(-0.8,1.5){\includegraphics{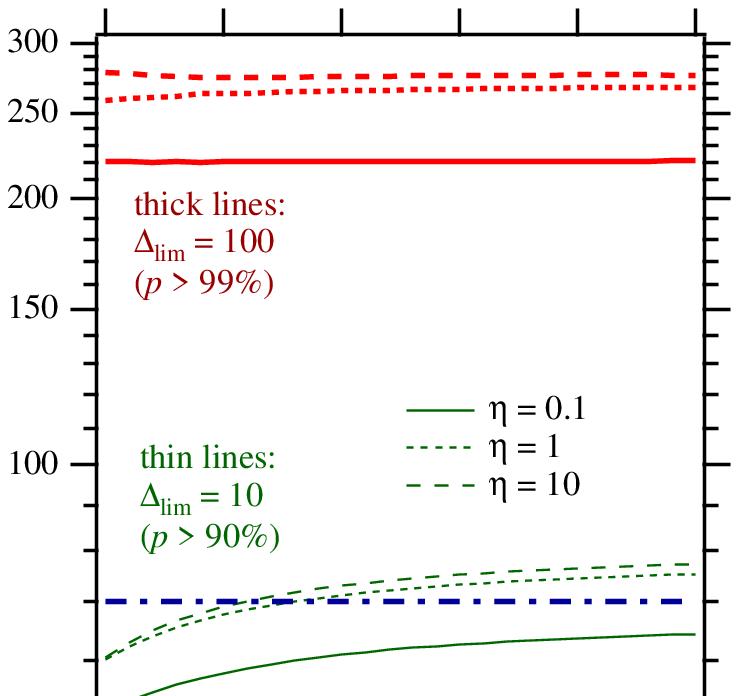}}  
 \put(8,1.5){\includegraphics{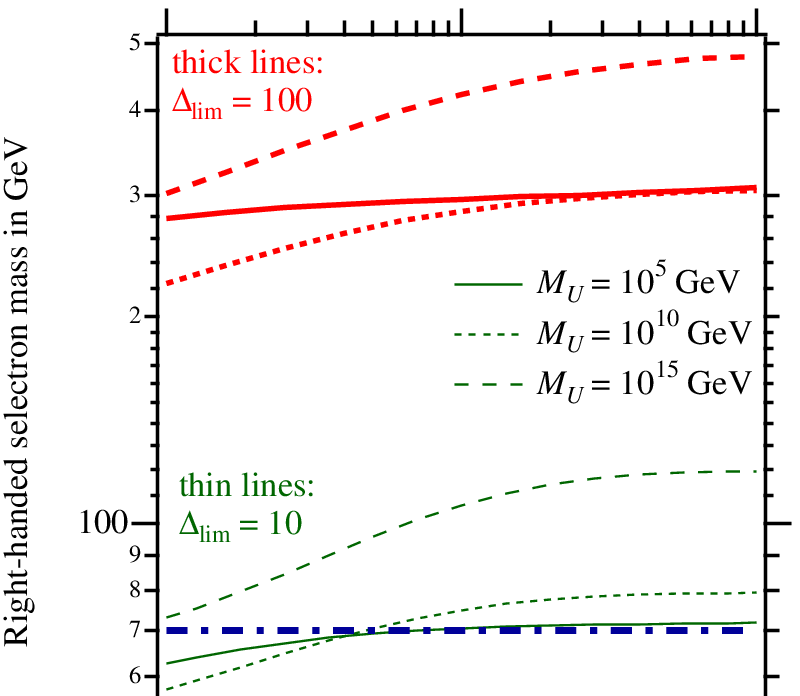}}}
\put(3.2,8.6){\ref{fig:2}a}
\put(12.4,8.6){\ref{fig:2}b}
\end{picture}
\caption[fig2]{\em The naturalness upper bound
$\Delta<100$ (thick lines) and $\Delta<10$ (thin lines)
on the right-handed selectron mass 
compared with its present experimental bound
(horizontal dot-dashed line) and plotted
as function of $\tan\beta$
for $M_U=10^6\GeV$ in figure \ref{fig:2}a, and
as function of $\eta$ for $\tan\beta=2$ in figure \ref{fig:2}b.
\label{fig:2}}
\end{center}\end{figure}

\begin{figure}[t]\setlength{\unitlength}{1cm}
\begin{center}\begin{picture}(15.5,8)
\ifMac
{\put(-1,0.5){\special{picture CP1}}
 \put(8,0.5){\special{picture CP2}}}
{\put(-1.2,0){\includegraphics{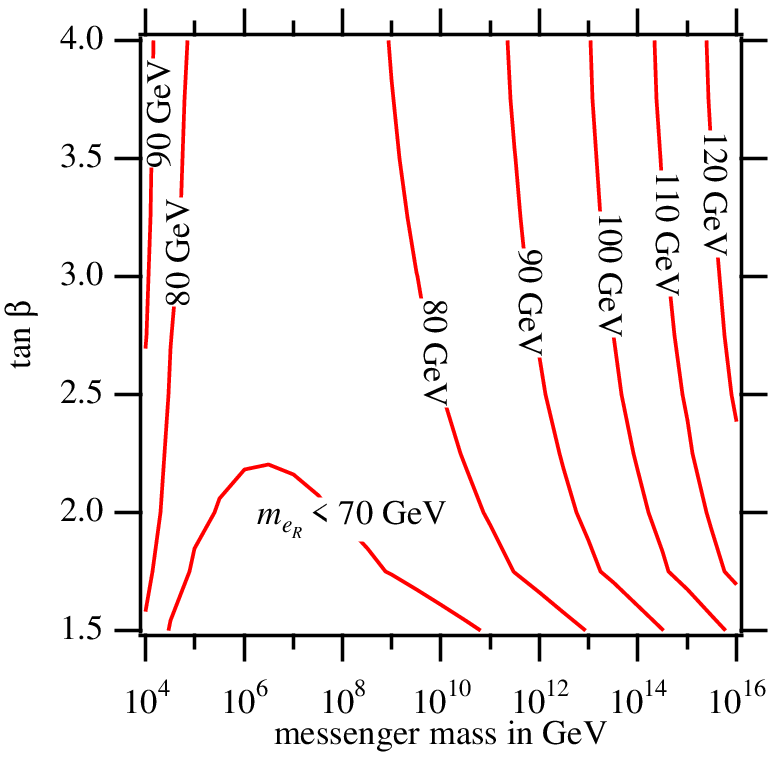}}  
 \put(8,0){\includegraphics{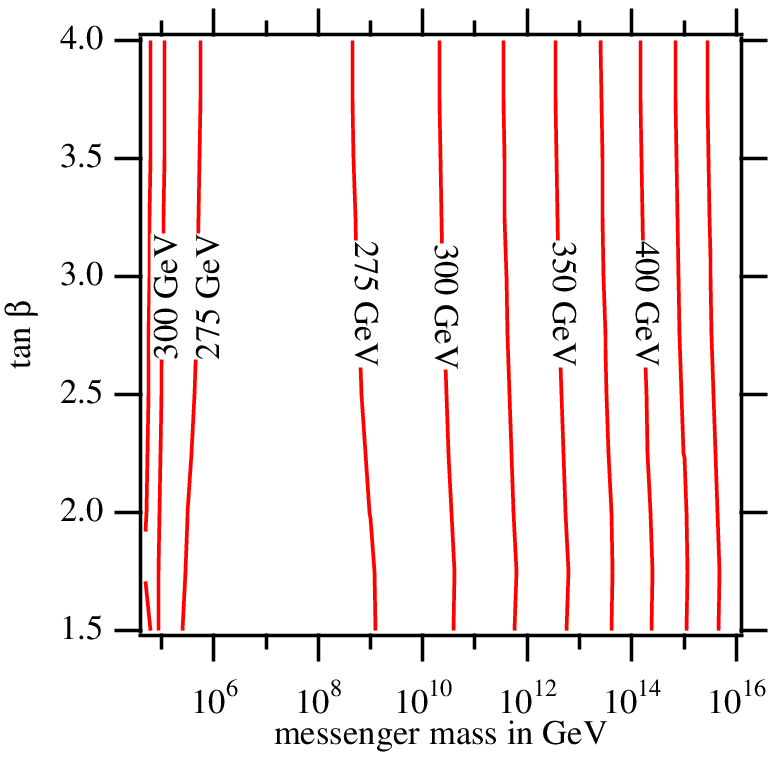}}}
\put(3,8.6){\ref{fig:CP}a}
\put(12.3,8.6){\ref{fig:CP}b}

\end{picture}
\caption[CP]{\em Contour-plot of the naturalness
upper bound
$\Delta<10$  (fig~{\rm \ref{fig:CP}a}) and
$\Delta<100$ (fig~{\rm \ref{fig:CP}b)}
on the right-handed slepton masses
as function of $(M_U,\tan\beta)$
in the minimal model with $\eta=1$.\label{fig:CP}}
\end{center}
\end{figure}

\section{Gauge-mediated soft terms}\label{GM}
In a large class of minimal models the soft terms are
predicted, at the messenger mass $M_U$, in terms of only
one scale, here called $M_0$:
\begin{equation}\label{eq:GM}
M_i(M_U)= \frac{\alpha_i(M_U)}{\alpha_{\rm GUT}} M_0,\qquad
m_R^2(M_U) =  \eta\cdot c^i_R M_i^2(M_U),\qquad
A^{\rm a}_g(M_U)=0,
\end{equation}
where $M_i$ are the three gaugino mass parameters,
$m_R^2$ are the soft mass terms for the fields
$R=\tilde{Q},\tilde{u}_R,\tilde{d}_R,\tilde{e}_R,
\tilde{L},h^{\rm u},h^{\rm d}$,
$g$ is a generation index and
${\rm a} = {\rm u,d,e}$.
The various quadratic Casimir coefficients $c^i_R$ are
listed in table~\ref{Tab:bcude}.
Finally, $\eta$ parametrizes the different minimal models.
For example
 $\eta=(n_5+3n_{10})^{-1/2}$ in models
with only one supersymmetry breaking singlet,
with $n_5$ copies of messenger fields in the $5\oplus\bar5$
representation of SU(5),
and $n_{10}$ copies in the $10\oplus\overline{10}$
representation~\cite{GMmu,GM96}.
Values of $\eta$ bigger than one are possible
if more than one supersymmetry-breaking singlet is present,
since an $R$-symmetry can suppress the gaugino masses
with respect to the scalar masses.

Before being able to compute the naturalness bounds
we must face the problem that gauge interactions alone cannot mediate
the (Peccei-Quinn breaking) `$\mu$-term',
as well as the corresponding `$B\cdot \mu$-term',
so that we control only a part of the contributions to
$M_Z^2$ in eq.\eq{MZ^2}.
Even worse, it is also possible that the unknown physics
required to solve the problem
gives rise to non minimal
contributions to the other soft terms,
more probably only to the ones in the Higgs sector~\cite{GMmu}.

If, in the worst case, also the sfermions
received unknown non-minimal contributions,
not only would it be impossible to convert the
upper bounds on the Higgs soft terms into upper bounds
on the sfermion masses
but also all the predictive power of gauge-mediation would be lost.
For this reason we do not study this possibility.

If instead unknown physics mediates new contributions
only to the Higgs soft terms,
the sfermion masses would be unaffected
apart from a RGE correction due to the hypercharge couplings.
This correction, that vanishes in the
non-minimal models of ref.~\cite{GMmu}, is non negligible only
if the masses of the Higgs fields $h_{\rm u}$ and $h_{\rm d}$ receive
largely different extra contributions.
Since such new contributions would enter additively
in the expression for $M_Z^2$
(at least until the relevant potential is the
RGE-improved MSSM one),
they would not affect the naturalness bounds on $M_0$.
This bound is, in fact, not evaded if unknown non
minimal contributions to $M_Z^2$ partially cancel
the minimal ones: we are
exactly going to compute how unnatural is this possibility.
We could, of course, parametrize the
unknown contributions to the Higgs masses
but the corresponding naturalness bounds would not give rise to interesting
upper bounds on the sfermion masses.

For these reasons we can limit ourselves to
compute the bounds on the minimal gauge-mediated contributions,
parametrized by $M_0$.
We can now begin our analysis observing that,
since the right-handed sleptons, $\tilde{e}_R$,
are the lightest sfermions\footnote{For moderate $\tan\beta$ the
right-handed sleptons are
degenerate. We will discuss in the following the case of large $\tan\beta$,
where a stau state becomes even lighter.}
the bounds on their masses are more interesting than
the bounds on the other heavier sfermions.
Secondly, this bound is strong, because $M_Z^2$,
as given in\eq{MZ^2}, receives
contributions proportional to the large squark masses
via the soft parameter $\mhu^2$ of the Higgs field $h_{\rm u}$
coupled to the top quark.

We can elucidate these points comparing this situation
with what happens in the other well motivated
case of supergravity mediated soft-terms.
Even making the strong and unmotivated assumption of universal
soft terms at the unification scale,
$m_R^2(\MGUT) = m_0^2$,
the naturalness bounds are not very interesting.
In fact, in the limit where the sparticle spectrum is dominated
by the contributions from $m_0^2$,
all the sparticle masses, and in particular
the Higgs mass parameters that enter the MSSM minimization
conditions, are of the same order, $m_R^2(M_Z)=\Ord(m_0^2)$
so that the naturalness bounds, computed in\eq{SuGraFT},
are not very interesting:
\begin{equation}
 m_{\tilde{e}} \approx
m_0 \circa{<} 600\GeV\sqrt{\frac{\Delta_{\rm lim}}{100}}
\end{equation}
On the contrary in minimal gauge-mediation models where soft mass terms are
generated by gauge interactions,
there is a hierarchy between the lighter sleptons,
that interact only weakly, and the heavier squarks and higgses,
with strong or $\lambda_t$ interactions (see eq.~(\ref{eq:GM})).
This hierarchy
enhances the naturalness bounds on the slepton masses.
This difference between the
two cases of supergravity and gauge-mediated soft terms
is exemplified in figure~\ref{fig:eRN},
where we plot the naturalness upper bounds
on the right-handed $\eR$ mass and on the
gaugino mass parameter $M_1$.
Note that in the unnatural regions we are exploring
$\mu$ is so large that $M_1$ coincides, approximately,
with the lightest neutralino mass.
We have chosen a small value of $M_U=10^5\GeV \ll \MGUT$:
for this reason the bound on the gaugino masses are weaker
than in the supergravity case, 
where the running starts from $\MGUT$.
Since the dependence on $\tan\beta$ is similar in the two cases,
we have fixed $\tan\beta=2$.

\smallskip

Having shown the interest of
the bounds on the $\eR$ masses,
we can now pass to a more detailed analysis.
The $Z$ boson mass can be easily determined as function of the supersymmetric parameters
by minimizing the RGE-improved tree-level potential,
that takes into account the dominant quantum effects,
generated at all the scales from $M_U$ down to $Q\approx M_Z$
and thus enhanced by large logarithms.
As is well known~\cite{V1loop}, in this approximation
the result has a strong unphysical dependence
on the choice of the low-energy scale $Q$.
This dependence can be reduced
using the better approximation of minimizing
the one-loop effective potential~\cite{V1loop,V1loopcalcolato},
that we have employed in producing the numerical results
plotted in the figures.
However, since the top/stop sector gives the dominant contribution
to the one-loop effective potential,
a good approximation 
is obtained making the stop corrections negligible
via an appropriate choice of the scale $Q\sim m_{\tilde{t}}$
and including only the top correction~\cite{V1loopcalcolato} to $M_Z^2$ in eq.\eq{MZ^2}
\begin{equation}
M_Z^2\to (1+\delta_{\rm top}) M_Z^2,\qquad\hbox{with}\quad
\delta_{\rm top} = \frac{1}{(4\pi)^2}\frac{\sin^2\beta}{1-\cot^2\beta}
\frac{8\lambda_t^4}{g_Y^2 + g_2^2} \ln \frac{m_{\tilde{t}}^2}{m_t^2}
\end{equation}
where $\lambda_t$ is the top quark Yukawa coupling in the MSSM.
Using now the solutions~(\ref{sys:RGEsol}) to the RGE given in appendix~A we get
\begin{equation}\label{eq:MZ^2(M0)}
(1+\delta_{\rm top}) M_Z^2 \approx -2\mu^2 + 1.2
\frac{(0.5+5\eta)+(0.1+2\eta) \cot^2\beta}{1-\cot^2\beta} M_0^2 +\cdots
\end{equation}
where we have fixed $M_U=10^5\GeV$. 
We remark that, in our approximation,
the possible non minimal additional contributions to the Higgs soft terms,
represented by the dots in eq.\eq{MZ^2(M0)},
enter additively even after RGE rescaling.
This does not happen in the one-loop quantum potential.
For this reason the full one-loop result has also a (negligible) dependence
on the remaining free parameters: the sign of $\mu$
and the possible non-minimal contributions to the Higgs soft masses.
In our plots we have set them to zero and chosen $\mu>0$.

For a typical value of $\delta_{\rm top} \approx 0.6$,
the naturalness bound on the
right-handed slepton masses is,
taking also into account the $D$-term contribution to the them
\begin{equation}\label{eq:meR<}
M_{\tilde{e}_R} \circa{<} 
M_Z\left(-\sin^2\theta_{\rm W}\cos 2\beta + \Delta_{\rm lim}
\frac{(\eta+0.025)(1-\cot^2\beta)}{(14\eta+1.5)+(5.5\eta+0.3)\cot^2\beta}\right)^{\!1/2}.
\end{equation}
We now discuss the dependence of this bound on the various parameters.
Expression\eq{meR<} says that, as usual,
the bound becomes stronger for small $\tan\beta$.
In the limit $\tan\beta\to 1$ the tree level Higgs potential
has no quartic couplings, so that the one loop corrections become important
and the full numerical result must be used.
We show this dependence in figure~\ref{fig:2}a for $M_U=10^6\GeV$
and for different minimal models.
The horizontal dot-dashed line represents the present
lower experimental bound on the $\eR$ mass.

The dependence on the parameter $\eta$ is
plotted in figure~\ref{fig:2}b.
We see that the bounds are stronger for smaller values
of $\eta$, obtained with a large messenger field content.
This is mainly a reflection of the fact that in such cases
the selectron is so light that it usually is the Lightest
Supersymmetric Particle (apart for the gravitino)~\cite{GM96}.
Unless the messenger mass is either near to the unification scale
or so small that the LSP decays within the detector,
and unless there are cancellations of more than one order of magnitude,
a large messenger field content is
disfavored by the experimental lower bound on the $\eR$ mass.

In figure~\ref{fig:CP} we finally show,
in the minimal model with $\eta=1$,
the dependence of the naturalness bounds
$\Delta_{\rm lim} < 10$ and $\Delta_{\rm lim} < 100$
on the messenger mass as a contour-plot in the plane $(M_U,\tan\beta)$.
We see that one-loop effects become more important in the case
of a more unnaturally splitted spectrum.
Due to the compensation of different effects
(higher $M_U$ means a longer running but also smaller squark masses
at $M_U$) the bounds are not much dependent on $M_U$.
For the lowest possible values of $M_U$ the running is so short that the
$\lambda_t$-induced negative contributions to $\mhu^2$ are just sufficient
to cancel the gauge-mediated contribution,
giving rise to much weaker naturalness bounds on the sfermion masses.
At the light of our naturalness considerations,
a fully natural acceptable spectrum is possible in this case.

In the same way we could obtain the corresponding
upper bounds on the masses of the other supersymmetric particles.
Since the typical correlations between different sparticle
masses have been carefully studied~\cite{GM96}
we do not need to report here the less interesting
bounds on the heavier sfermions, even if they could be
more stable with respect to non minimal corrections.
We also avoid discussing the bounds on the Higgs masses
because, since they contribute directly to $M_Z^2$ in eq.\eq{MZ^2},
these bounds are not characteristic
of the specific model of soft terms we are considering.

Before concluding, we observe that in the case of large
$\tan\beta\sim m_t/m_b$,
a stau becomes significantly lighter than
the other sleptons.
This is due both to the renormalization corrections
induced by $\lambda_\tau$, and to the 
left/right mixing term in the stau mass matrix.
Since, especially for $\Delta<100$, the stau masses depend in
an important way also on the unknown $\mu$ term, we limit
ourselves to observe that the naturalness bound
on the lightest $\tilde{\tau}$ mass is typically
two times stronger than the ones we have so far presented on the
$\tilde{e}_R$ mass.
As usual~\cite{LargeTan}, additional cancellations may be necessary to
satisfy the other minimization condition that should give a large
value of $\tan\beta$.

\smallskip

Finally, as an aside remark,
we point out that
--- again because the Higgs soft term $\mhu^2$
is large and negative --- 
unless there are large non-minimal corrections
to the soft terms, when
$M_U\circa{>}10\TeV/\lambda_{b,\tau}$
the combination 
$m_2^2\equiv \mhu^2+m_{\tilde{L}}^2$ turns out to be negative.
In this case the SM-like minimum is not
the unique vacuum state:
the MSSM potential develops other
deeper unphysical minima
with $\md{h_{\rm u}}\approx \md{\tilde{L}}$~\cite{L<hu}.
The value of the fields at the minimum is given by the
scale $Q_0$ at which $m_2^2(Q_0)=0$, and is typically
one or two orders of magnitude below the mediation scale $M_U$.
Alternatively, if $Q_0$ is sufficiently large, non renormalizable
operators (for example those responsible of the neutrino masses)
can stabilize the potential along the direction $ \tilde{L}h_{\rm u}$
at scales smaller than $Q_0$.

However this kind of unphysical minimum is not a problem, since,
as we will now discuss,
the evolution of the universe will likely prefer the SM-like minimum.

First of all, the only minimum present during the
high-temperature phase is the symmetric one, $\phi=0$.
If the relevant fields $\phi = \{h_{\rm u},\tilde{L},\tilde{Q},\ldots\}$
are not displaced from it when the universe
cools down below the Fermi scale,
the electroweak phase transition~\cite{Thermal}
results in the usual SM-like vacuum.
If, on the contrary, the relevant fields
have a large vacuum expectation value
$\phi_R$ at the end of inflation
(as would happen if the unphysical minimum were present
also during inflation\footnote{A non zero $\phi_R$ after inflation
can be pushed to the origin $\phi=0$ during the pre-heating stage
in models where the inflation field
decays resonantly to a field 
coupled to the flat direction
$\tilde{L}h_{\rm u}$~\cite{PreRiscaldamento}.}),
they will rapidly oscillate around the symmetric
minimum with amplitude decreasing as
$\phi(T) = \phi_R\times (T/T_R)$ where
$T_R$ is the reheating temperature\footnote{We are interested in the case
where $\phi\circa{>} T$, so that the fermions coupled to $\phi$ receive
large masses that prevent its decay.}.
If $T_R>\phi_R$ thermal effects are therefore sufficient to solve
the cosmological problem posed by the unphysical minimum.

Fortunately, the quantum corrections that give rise
to the unphysical minima are not present
during inflation~\cite{Infl}.
For this reason we expect that $\phi_R=0$ so that the SM-like minimum
is preferred by cosmological history.
To conclude that the situation is absolutely safe,
we must ensure that the
the tunneling rate towards the unphysical minimum is small enough.
The rate, computed in~\cite{Infl},
is in fact suppressed by a semi-classical
factor $\exp\Ord(-1/\lambda_{b,\tau}^2)$
and turns out to be small enough even if $\tan\beta$ is large,
unless the $\mu$-term is very small.


\section{The NMSSM with gauge-mediated soft terms}\label{NMSSM}
An elegant attempt to overcome the difficulty of generating the $\mu$-term
consists in modifying the field content of the MSSM
in such a way that the $\mu$-term arises through the vacuum expectation
value of a singlet $S$ with superpotential obtained replacing~\cite{NMSSM}
\begin{equation}
\mu h^{\rm u} h^{\rm d}\quad\to\quad \lambda\,
S h^{\rm u} h^{\rm d} - \frac{\kappa}{3} S^3
\end{equation}
so that $\lambda\md{S}=\mu$.
The `$B\cdot\mu$' term is also naturally generated.

This model (`NMSSM') has been extensively studied in the
case of supergravity-mediated soft terms~\cite{NMSSMSuGra}
finding that
there are often inequivalent physically acceptable
minima with different values of $\md{S}$,
sometimes together with unphysical minima where,
for example $\md{S}=0$.
We point out that, in such cases,
considering the
thermal corrections to the potential and
discussing the electroweak phase transition, for example along
the lines of ref.~\cite{Thermal},
it is possible to compute which one
of the various minima --- not necessarily the deepest one ---
is selected by cosmological evolution.
For example, if $\mhd^2,m_S^2>0$ and $\mhu^2<0$ the chosen minimum will be
the one that at zero temperature
is not separated by a potential barrier from the symmetric state where
$S=h_{\rm u}=h_{\rm d}=0$.
This property makes simple to find it numerically.
If deeper minima are present, 
it is also necessary to check if the tunneling rate
is small enough.
If instead $m_S^2$ is negative there are two different
such minima and identifying the selected one
becomes a very cumbersome numerical problem.

These considerations are superfluous
in the case of gauge-mediated soft terms,
where the only minimum of the NMSSM potential is
the unphysical one
$$\md{h_{\rm u}}= \frac{2|\mhu|}{(g_Y^2+g_2^2)^{1/2}},\qquad
\md{h_{\rm d}}=\md{S}=0.$$
The reason is that
the soft parameters of the singlet field,
$m_S^2$, $A_\kappa$ and $A_\lambda$,
zero at tree level, remain too small
with respect to $\mhu$~\cite{Dine,GMmu}.
The only possibility of avoiding this unfortunate situation
consists in having a light and fine-tuned messenger spectrum
such that
$\mhu^2(Q)\sim m_S^2$.
In this case $M_Z^2=\Ord(m_S^2)$ so that this fine-tuning
also guarantee heavy sfermions and gauginos.
Note also that in such a case $\tan\beta$,
as obtained from the NMSSM minimization condition,
$$\sin 2\beta=\frac{2\kappa \lambda S^2-A_\lambda\lambda S}
{\mhu^2 + \mhd^2+\lambda^2 (2S^2 + v^2)}\sim 
\frac{m_S^2}{\mhd^2(\tan\beta)},$$
is large, $\tan\beta\sim 50$.
When $m_S^2$ is not much smaller than $1/50$ of the
`natural' value of the Higgs soft terms, $\sim\mhu^2(M_U)$,
the necessary fine tuning is not much worse than the
minimal one, $\Delta\circa{>} 50$,
anyway necessary to obtain a large $\tan\beta$~\cite{LargeTan}.

\begin{table}
$$\begin{array}{c|ccccc|ccc||c||ccc}
\phantom{-}b_i&c_i^Q&c_i^{u_R}&c_i^{d_R}&c_i^L&c_i^{e_R}&
c_i^{\rm u}&c_i^{\rm d}&c_i^{\rm e}&i,g&
b^{\rm u}_g&b^{\rm d}_g&b^{\rm e}_g\\[0.5mm] \hline
\phantom{-}{33\over5} &{1\over30}&{8\over15}&{2\over15}&\vphantom{X^{X^X}}
{3\over10}&{6\over5}& {13\over15} &{7\over15}&{9\over5}&1&3&0&0\\
\phantom{-}1&{3\over2}&0&0&{3\over2}&0&3&3&3&2&3&0&0\\
-3&{8\over3}&{8\over3}&{8\over3}&0&0&{16\over3}&{16\over3}&0&3&6&3&0
\end{array}$$
\caption{\em Values of the RGE coefficients in the MSSM.
The coefficients $c^{h_{\rm u}}_i$ and $c^{h_{\rm d}}_i$
are equal to $c^L_i$.
\label{Tab:bcude}}
\end{table}


\appendix
\section{Renormalization of gauge-mediated soft terms}
Neglecting all couplings except the gauge and the top Yukawa ones,
the solutions to the one loop RGEs between
the messenger mass $M_U$ and $M_Z$
may be written in terms of analytic
functions and only one
function, $\lambda^{\rm max}_t(E)$,
calculable only numerically~\cite{RGEinMSSM}.
We abbreviate the various functions of $E$, $\varphi(E)$, as $\varphi_E$ and
define $t_E\equiv (4\pi)^{-2}\ln\MGUT^2/E^2$.

For simplicity we assume,
consistently with their measured values,
that the gauge couplings satisfy unification relations so that we can
parametrize $\alpha_i$ ($i=1,2,3$) and $\lambda_t$ in terms of
$\alpha_{\rm GUT}$ and $\lambda_{t\rm GUT}$ in the usual way
\begin{eqnsystem}{sys:MSSMsol0}
\alpha_i(E) &=& \alpha_{\rm GUT}/[1+b_i g_{\rm GUT}^2 t(E)], \\
\rho_E\equiv\frac{\lambda_t^2(E)}{\lambda^{\rm 2max}_t(E)} &=&
[1+\lambda^{\rm 2max}_t(E)/\lambda_{t\rm GUT}^2 E_{\rm u}(E)]^{-1},
\end{eqnsystem}
where we have defined
$$
E_\alpha(E)\equiv\prod_i
\left[\frac{\alpha_{\rm GUT}}{\alpha_i(E)}\right]^{c_i^\alpha/b_i},
\qquad
\lambda^{\rm max}_t(E)=\Bigg[
 2b^{\rm u}_3 \int^{\ln\MGUT}_{\ln E}\!
\frac{E_{\rm u}(E')}{E_{\rm u}(E)} d\ln E'  \Bigg]^{-1/2}.
$$
The $\beta$-function coefficients $b_i$, listed in table~\ref{Tab:bcude},
include only the contribution from the MSSM fields.
Consequently
$\alpha_{\rm GUT}\approx 1/24$ and $\lambda_{t\rm GUT}$
are then just parameters in terms of which it is convenient
to express the solutions of the RGE below $M_U$ for
the gaugino masses $M_i$, for
the trilinear terms $A^{\rm a}_g$ (where $\rm a = u,d,e$ and
$g=1,2,3$ is a generation index)
and for the soft masses $m_R^2$ of the fields $R$.
The values of these parameters at the Fermi scale
corresponding to the boundary
conditions\eq{GM} are
\begin{eqnsystem}{sys:RGEsol}
M_i &=& M_0\cdot \alpha_i/\alpha_{\rm GUT}\\
A^{\rm a}_g &=&  (x_1^{\rm a} + {\textstyle\frac{1}{6}}b^{\rm a}_g I')M_0\\
m_R^2 &=& m_R^2(M_U)+x_2^R M_0^2.\\[3mm]
\noalign{\hbox{The soft masses of the
fields involved in the top quark Yukawa coupling
get the additional corrections}}\nonumber\\[-3mm]
m_{h_{\rm u}}^2  &=& m_{h_{\rm u}}^2(M_U)+x_2^{h_{\rm u}} M_0^2-{1\over2}I\\
m_{\tilde Q_3}^2 &=& m_{\tilde Q_3}^2(M_U)+x_2^Q M_0^2-{1\over6}I\\
m_{\tilde t_R}^2 &=& m_{\tilde t_R}^2(M_U)+x_2^{u_R} M_0^2-{1\over3}I
\end{eqnsystem}
where we have defined
$$\begin{array}{ll}\displaystyle
x_{nU}^R \equiv\sum_{i=1}^3{c_i^R\over b_i}
[1-\frac{\alpha_i^n(M_U)}{\alpha^n_{\rm GUT}}], &
\phi_E \equiv \rho_E-6\lambda_{tE}^2 t_E,\\[2mm]
\displaystyle
x_{n\phantom{U}}^R \equiv\sum_{i=1}^3{c_i^R\over b_i}
[\frac{\alpha_i^n(M_U)}{\alpha^n_{\rm GUT}}-
\frac{\alpha_i^n(M_Z)}{\alpha^n_{\rm GUT}}],\qquad &
\Phi_E \equiv \phi_E^2 - 6\lambda_{tE}^2 t_E^2 (c^{\rm u}_i g_i^2)_E,
\end{array}$$
and
\begin{eqnsystem}{sys:PantaMSSM}
I &\equiv&
[m_{h_{\rm u}}^2(M_U)+
m_{\tilde Q_3}^2(M_U)+ m_{\tilde t_R}^2(M_U)]\cdot
\frac{\rho_Z-\rho_U}{1-\rho_U} + M_0^2\cdot\left[
\frac{(1-\rho_Z)(\rho_Z-\rho_U)}{(1-\rho_U)^2}(\phi_U+x_{1U}^{\rm u})^2+
\right.\nonumber \\ &&\left.\nonumber
+2\frac{1-\rho_Z}{1-\rho_U}(\phi_Z-\phi_U)(\phi_U+x_{1U}^{\rm u})
-(\Phi_Z-\Phi_U)
-\frac{\rho_Z-\rho_U}{1-\rho_U}(\Phi_U+x_{2U}^{\rm u})\right]
\\ \nonumber
I' &\equiv&(\phi_Z-\phi_U) +
\frac{\rho_Z-\rho_U}{1-\rho_U}
(x_{1U}^{\rm u}+\phi_U)
\end{eqnsystem}
All the coefficients are listed in table~\ref{Tab:bcude}.



\end{document}